\documentclass[preprint,amsmath,amssymb,groupedaddress]{revtex4-1}
\usepackage{graphicx}% Include figure files
\usepackage{epstopdf}
\usepackage{dcolumn}% Align table columns on decimal point
\usepackage{bm}% bold math
\usepackage{multirow}
 \usepackage{slashbox}
\begin{document}
\preprint{In press in JCP, Volume 141, Issue 2, page x (2014)}
\title{Influence of surface coverage on the chemical desorption process}% Force line breaks with \\
   \author{M.~Minissale}
\email{email adress: marco.minissale@obspm.fr}
\author{F. Dulieu}%
\email{email adress: francois.dulieu@obspm.fr}
\affiliation{%
Universit\'e de Cergy Pontoise and Observatoire de Paris, ENS, UPMC,
UMR 8112 du CNRS\\ 5, mail Gay Lussac, 95000 Cergy Pontoise cedex,
France.}
\begin{abstract}
In cold astrophysical environments, some molecules are observed in the gas phase whereas they should have been depleted, frozen on dust grains. In order to solve this problem, astrochemists have proposed that a fraction of molecules synthesized on the surface of dust grains could desorb just after their formation. Recently the chemical desorption process has been demonstrated experimentally, but the key parameters at play have not yet been fully understood. In this article we propose a new procedure to analyze the ratio of di-oxygen and ozone synthesized  after O atoms adsorption on oxidized graphite.  We demonstrate that the chemical desorption efficiency of the two reaction paths (O+O and O+O$_2$) is different by one order of magnitude. We show the importance of the surface coverage: for the O+O reaction, the chemical desorption efficiency is close to 80 $\%$ at zero coverage and tends to zero at one monolayer coverage. The coverage dependence of O+O chemical desorption is proved by varying the amount of pre-adsorbed N$_2$ on the substrate from 0 to 1.5 ML. Finally, we discuss the relevance of the different physical parameters that could play a role in the chemical desorption process: binding energy, enthalpy of formation, and energy transfer from the new molecule to the surface or to other adsorbates.\end{abstract}
\pacs{}% PACS, the Physics and Astronomy
\keywords{Surface reactions, Chemical desorption, Coverage, Energy transfer, Oxygen}

\maketitle
%\tableofcontents
\section{Introduction}
On cold surfaces (T$<$10K) most species can adsorb in physisorbed states and remain trapped \cite{Collings04}.
If the species are reactive, some chemical transformation can occur, without help of any external agents like photons, electrons or other energetic particles. Thanks to these trapping properties, cold dust particles are the nano-catalysts of cold environments. Since decades, a large part of the molecular diversity observed in our Universe is attributed to the catalytic role of interstellar dust.\cite{Oort46} The dust can be made from carbonaceous materials, silicates and be coated with water. It is usually considered to have a temperature T$<$20 K in most of the environments of the interstellar medium. 
Radio-astronomy has powerful eyes able to decipher our ancient and primordial chemical history, thanks to the freeing of rotation in the gas phase. Unfortunately, the birth-place of the molecular complexity is partially hidden and locked in the solid phase. For this reason, freezing and synthesis of complex molecules on micron-sized cold dust particles remain ambiguous: solid and gas phase equilibrium is still an open problem. Pagani et al\cite{Pagani12} have shown that CO and N$_2$ species present very different depletion behaviors, in spite of their similar sticking probability and desorption efficiency.\cite{Bisschop06}
Bacmann et al\cite{Bacmann12} have observed unexpected ``high'' gas-phase abundances of large molecules like CH$_3$OCH$_3$ in very cold environments (i.e. pre-stellar cores): such molecules should remain trapped on the solid phase. 
This is a direct (or paradoxal) observational evidence of the link between solid and gas phase. 
Takahashi and Williams \cite{Takahashi00} proposed that desorption of CO is induced by chemistry; the energy released by a H+H reaction heats the grain and provokes the desorption of CO from very small grains. More recently, different authors propose direct reactive desorption mechanisms.\cite{Garrod07,Cazaux10} For example Vasyunin and Herbst\cite{Vasyunin13} include in their model a reactive desorption parameter: the efficiency of the chemical desorption process is constrained to 1-10$\%$ in order to reproduce the astrophysical observations.

From a theoretical point of view, desorption upon formation has been linked to indirect mechanisms, as in the case of photo-dissociation of water,\cite{Andersson06} or to direct formation mechanisms, like in the case of H+H or O+H.\cite{Morisset04, Bergeron08}  

Photo-desorption experiments have confirmed that recombinative desorption is a possible mechanism,\cite{DeSimone13} especially in the case of irradiation of O$_2$ and CO$_2$.\cite{Fayolle13, Fillion14} Moreover, Rajapan et al.\cite{Rapajan11} have shown that surface plays a fundamental part in the total efficiency of the process. The presence of a SiO$_2$ surface (as a third body) enhances the rate of combination of radical species (NO$_2$ and N$_2$O$_4$) produced by photolysis, compared to the gas phase. Nevertheless photo-desorption has a limited efficiency of about 1 desorption per 1000 incident photons. Moreover it requires usually more energy (typically 7-10 eV) than those of chemical reactions ($<$ 4 eV) and even more compared to binding energies ($<$ 0.4 eV). On the contrary, Dulieu et al,\cite{Dulieu13} by experimentally studying water formation, show that direct chemical desorption can have an efficiency close to unity. In two previous works,\cite{Chaabouni12, Dulieu13} we showed that the chemical desorption efficiency depends on the substrate, and it is maximized for bare surfaces. 
Furthermore, we demonstrated that the type of reaction affects chemical desorption efficiency: for example OH+H $\rightarrow \rm{H_2O}$ reaction has a very high probability of chemical desorption, while some others, like O$_2$H+H$\rightarrow \rm{H_2O_2}$, appear to not undergo chemical desorption. So far no real clues have been proposed yet to understand such differences.

The chemical desorption process starts from the energy excess usually present in radical-radical reactions. Chemical desorption efficiency will depend on how the newly formed molecule dissipates the energy excess. Actually, in order to desorb, the molecule has to convert a fraction of this excess formation energy into kinetic energy, and especially into motion perpendicular to the substrate. 
In other words, the problem lies on how the newly formed molecule manages the energy excess and interacts dynamically with the surface. The total budget of energy excess (enthalpy of reaction) is the most important parameter describing the chemical desorption. The larger is the enthalpy of reaction, the more probable is chemical desorption. In addition, the binding energy of the newly formed species should be considered as a limiting factor. 
Among different parameters, the degrees of freedom of newly formed molecules seem to play a important role: in fact more degrees of freedom lead to an easier distribution to internal modes and thus dissipation. 
The last parameter that we consider is the interaction with the substrate, namely the phonon propagation. Water ice substrate has larger capabilities to dissipate the excess of energy than other substrates (silicates and oxidized graphite), and therefore the probability of chemical desorption is lower. Nevertheless, the ``softness'' of water ice substrate is not sufficient to forbid chemical desorption, as shown in Amiaud et al\cite{Amiaud07} and Govers et al\cite{Govers80} in the case of H$_2$. Further works stress - in the case of water - the role of the surface coverage in the inhibition of the chemical desorption process.\cite{Congiu09}

The aim of this article is to describe the role and the contribution of the different parameters influencing the chemical desorption. We have chosen to present the case of O atom reactivity on oxidized graphite for two reasons. 
First, O atoms reactivity is a well known system,\cite{PriceX, Minissale13, Minissale14, Congiu14} and, since only 2 reactions are involved:
%\begin{align}
$$\rm{O + O} \longrightarrow  O_2 \qquad (R1) $$  %\\
$$\rm{O + O_2} \longrightarrow  O_3\qquad  (R2),$$ %\\
%\end{align}
we can immediately realize how the different parameters affect the two reactions.  Second, among substrates already studied, the oxidized graphite exhibits the highest chemical desorption efficiency. 

The paper is organized as follow: Sec.~\ref{sec:experiments} presents briefly the experimental protocol and results. In Sec.~III, we analyze experimental results and we present a model to simulate chemical desorption and constrain the above-mentioned parameters. In Sec.~IV, we discuss the physical meaning of these findings. In Sec.~V, we discuss the main conclusions of this work.

\section{Experimental and results}\label{sec:experiments}

Experiments have been performed using the FORMOLISM set-up.\cite{Amiaud08,Congiu12} The experiments described here are very similar to those carefully detailed in Minissale et al,\cite{Minissale14} except for the substrate used. Experiments take place in an ultra-high vacuum chamber (base pressure 10$^{-10}$ mbar), containing the HOPG sample  (0.9~cm in diameter) stuck on the copper finger of the cryostat, operating at temperatures between 6.5~K and 350~K.  

The temperature of the sample (T$_s$) is computer-controlled, by a calibrated Silicon-diode and a thermocouple (AuFe/Chrome l K-type) clamped on the sample holder. The effective surface temperature is checked by using adsorption-desorption cycles of multilayer of different species (D$_2$, O$_2$, CO$_2$, H$_2$O). Before performing the experiments presented here, the HOPG sample has been oxidized through exposure to an O-atom beam. The oxidation is deduced from the irreversible modification of the Thermally Programmed Desorption (TPD) profiles of different adsorbates. Experiments are fully reproducible once the initial oxidation step is achieved.
Via a triply differentially pumped and collimated beam , O atoms and O$_2$ molecules in the ground state ($^3$P and X$^3\Sigma_g$ respectively) are aimed to achieve a uniform distribution at the center of the cold (6.5-25~K) sample, into a $\approx$ 3$\times$4~mm elliptical zone. O atoms are produced from O$_2$ cracking and typical value of $\tau$~= 70~$\pm$~5\% is used for the dissociation fraction. The flux $\Phi$~= 6$\pm$2$~\times$~10$^{12}$ O atoms/cm$^2$ (under O or O$_2$ form) is calibrated by looking at the appearance of O$_2$ multilayer features in TPD for different beam exposure times, the dose. The unit used in this article is ML, which is a surface density unit of 10$^{15}$ molecules per cm$^2$.
The products are probed using TPD and Reflexion Absorption Infrared Spectroscopy (RAIRS), even if in this paper we will not present infrared data. The RAIRS spectra confirm that all the reactions are taking place during the exposure of the atoms. The TPDs are performed by increasing the temperature at 10 K/min, and are used here because of their excellent signal to noise ratio. Most of the uncertainties (around 5~\%) stems from the control of the deposited doses.\\

\begin{figure}
   \centering
      \includegraphics[width=8.6cm]{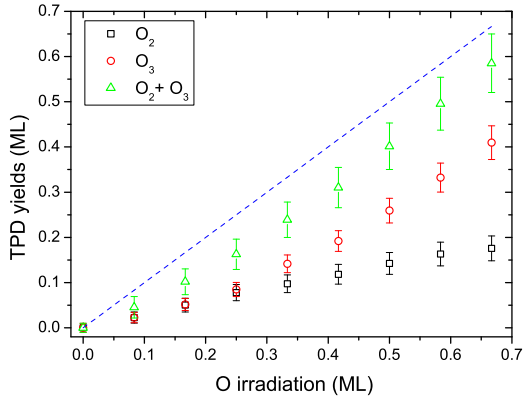}
   \caption{Integrated TPD spectra yields of O$_2$ (squares) and O$_3$ (circles) obtained after exposition of different doses of O (75~\%) and O$_2$ (25~\%). The straight line corresponds to a full conservation of the O atoms.}
              \label{fig:1}
    \end{figure}
%_________________________________________________________________

\begin{figure}
 \centering
  \includegraphics[width=8.6cm]{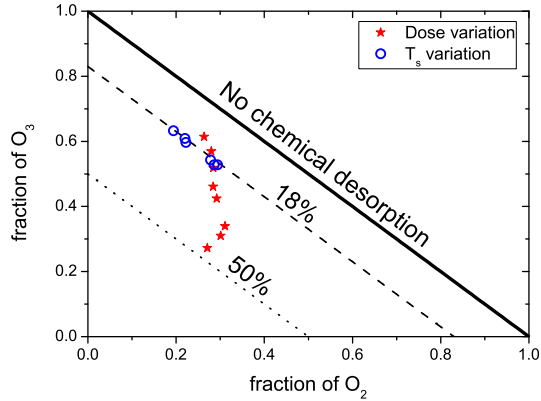}
   \caption{Fraction of O$_2$ and O$_3$. Red stars correspond to experiments shown in Fig.~\ref{fig:1} (variation of the dose at fixed surface temperature, 10~K), and blue circles correspond to different depositions of fixed dose (0.5 ML) at different surface temperatures (8, 10, 15, 20, 22, and 25~K). Solid, dashed and dotted lines are obtained using $f_{CD}$=0, 18, and 50~\%, respectively.}\label{fig:2}
    \end{figure}

Fig~\ref{fig:1} shows the results of a first set of experiments. We depose on the sample held at 10~K different doses (0, 0.08, 0.17, 0.25, 0.33, 0.42, 0.5, 0.58 , and 0.67~ML respectively) of O/O$_2$ mixture from the dissociated beam.  O$_2$ (squares) and O$_3$ (circles) yields (and their sum, triangles) are observed at the end of the exposure via integration of TPD traces. Thanks to its fast diffusion (about 10$^{-16}$cm$^2$/s at 10~K),\cite{Congiu14} O atoms can react with themselves (R1) to form O$_2$ or react with O$_2$ (R2) to form O$_3$. The balance between (R1) and (R2) is dependent on the coverage. In Fig.~\ref{fig:1}, we choose to multiply the O$_3$ signal by 3/2, in order to represent twice the total number of O atoms. The blue straight line corresponds to the conservation of the total number of O atoms: if all the atoms (in the form of O, O$_2$, and O$_3$) desorb during the TPD, therefore the triangles should overlie this line. This line is made experimentally using the undissociated beam (pure O$_2$ beam). We did not detect any signal that could be interpreted as O-atom desorption in the gas phase during the TPD, and the sum O$_2$+O$_3$ does not fit the blue line, in contrast to what is observed on a water substrate.\cite{Minissale14} The large deficit in O atoms is due to the reactive desorption or ``chemical desorption''. The excess of chemical energy of the newly formed molecule is partly transferred to kinetic energy  (by bounce on the substrate), and desorption may occur. We can also remark on Fig.~\ref{fig:1} that the O$_2$/O$_3$ ratio depends on the coverage, or more precisely, that the fraction of missing O is reduced with the coverage.\\
To get a better evidence of the change in chemical desorption efficiency, we re-plot in Fig.~\ref{fig:2} the results of Fig.~\ref{fig:1}. We use as x-axis the fraction of O$_2$ ($f$O$_2$) and as y-axis the fraction of O$_3$ ($f$O$_3$). $f$O$_2$ is obtained by dividing the yield of O$_2$ ($Y$O$_2$) by the dose $D_f$ (defined  as the flux $\Phi$ multiplied by the exposure time $t_f$), and the fraction of O$_3$  is calculated the same way ($f$O$_3 =\frac{YO_3}{D_f}$).
We can define the efficiency of chemical desorption as $f_{CD}$ =1 - $f$O$_3$-$f$O$_2$. The equality $f$O$_3$+$f$O$_2$=1 means that all atoms have remained on the surface before the TPD and so the chemical desorption is fully inefficient. We have represented this case using a broad black line on The Fig.~\ref{fig:2}. On the contrary, if there is no O$_2$ or O$_3$ desorption during TPD, all the atoms should have gone, $f_{CD}$ =1, and the chemical desorption is fully efficient. 

Fig.~\ref{fig:2} shows that $f_{CD}$ changes for experiments where dose has been changed (red stars, same data as in Fig.~\ref{fig:1}). The lowest coverage lies just above the dotted line which corresponds to 50~$\%$ of chemical desorption, while the highest coverage lies above the 18~$\%$ chemical desorption line. With the increase of the coverage, there is a rather vertical shift up. This shift illustrates two facts: (i) the O$_2$ production saturates because O$_2$ molecules are transformed into O$_3$ molecules by new incoming O atoms; (ii) the larger the coverage, the lower the chemical desorption efficiency. 
We want to stress that for increasing doses we need to increase exposure times, keeping the flux of the atomic beam constant. The measurement is made at the end of the exposure, but the coverage evolves during the exposure phase. An experimental point does not represent a constant coverage, but the evolution from zero to the final coverage. Actually, our experiments represent integration over the time of events. From the stars in Fig.~\ref{fig:2}, it is clear that chemical desorption efficiency is reduced with the coverage, but the used experimental method partly conceals the correlation between the coverage and the chemical desorption, because the coverage is not the parameter directly  varied. 
For this reason we perform a second set of experiments. This time, while keeping constant the dose (0.5~ML) - using the same evolution of the coverage (at first approximation) -  we increment the surface temperature for different experiments from 8 to 25~K (above which O$_2$ starts to desorb). A change of the surface temperature varies the diffusion properties, and as a consequence, changes the balance of two reactions (R1) and (R2) and the final products. These experiments are represented with blue circles in Fig.~\ref{fig:2}. All the points are located around the 18$\%$ dashed line. The O$_2$/O$_3$ balance changes. The higher is the surface temperature, the larger is $f$O$_3$, as expected. More surprisingly, the chemical desorption seems to be rather insensitive to the final product ratio.
 
\section{Analysis}

  \begin{figure}
 \centering
  \includegraphics[width=8.6cm]{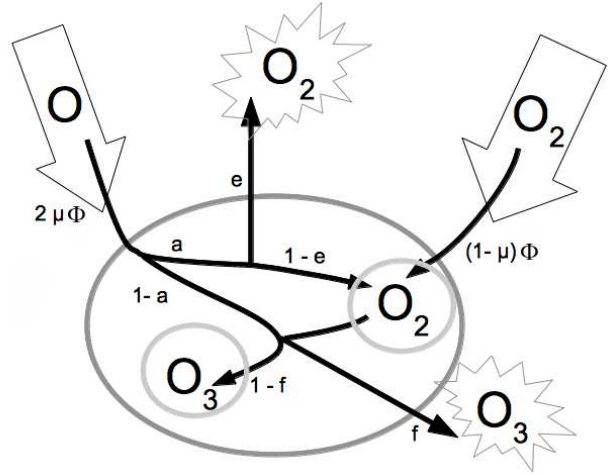}
   \caption{Schematic view of the processes occurring on the surface. $e$ and $f$ are the chemical desorption probability of (R1) and (R2) respectively.}\label{fig:3}
    \end{figure}

To analyze our experimental result, we can consider how the populations of species adsorbed on the substrate evolve with time (or coverage). In particular, there are three populations evolving with the deposited dose: O(t), O$_2$(t), and O$_3$(t). However, we are only able to measure O$_2$( t$_f$) and O$_3$(t$_f$) final populations, represented through grey circles on Fig.~\ref{fig:3}. The evolution of populations has two origins: external incoming fluxes of O and O$_2$ molecules (represented by broad arrows), and chemical evolution of the populations. O atoms undergo the two reactions (R1) and (R2). The branching ratio between (R1) and (R2) is expressed by the parameter $a$. If $a$=1, only (R1) takes place, if $a$=0, only (R2) takes place. Clearly the balance between the two reactions is dependent on populations, so from one experiment to another, and even during any given experiment, $a$ varies ranging always between 0 and 1. The parameter $e$ describes the chemical desorption of (R1), and the parameter $f$ describes the chemical desorption of (R2). If $e$=1, all the O$_2$ formed via O+O reactions steadily desorb. Parameters $e$ and $f$ actually describe the chemical desorption process. 

We now write the O$_2$(t$_f$) population, composed of the three following parts:
\begin{itemize}
\item[1] The contribution from the undissociated part of the beam: $$(1-\mu)\, \Phi\, t_f $$
\item[2] The contribution of (R1) (O part of the beam) less the part that has undertaken chemical desorption: $$\mu\, \Phi\, t_f\,  a\, (1-e)$$
\item[3] The subtraction due to (R2). Here we can remark that the reduction in O$_2$ is made by reaction from O atoms that have not been used in (R1). By using this trick, we can use the $a$ parameter without having to add a new parameter. Thus the negative contribution of (R2) to O$_2$ population is: $$ - 2 \, \mu\, \Phi\, (1-a) $$
\end {itemize}
By adding the three contributions, we obtain:
$$O_2\,(t_f)= \mu \, \Phi \, [t_f \, (\frac{1}{\mu} + a - a\,e - 1) + 2 \,(1-a)].$$
Similarly, we can write the O$_3$(t$_f$) population: $$O_3\,(t_f)= 2\,\mu\,\Phi \,(1-a)\, (1-f).$$
 
We note here that we have two observables, O$_2$(t$_f$) and O$_3$(t$_f$), and three parameters, $a$, $e$, and $f$. $a$ is highly variable and depends on experimental conditions, and especially surface temperature. 
By constructing a parametric plot of all possible solutions for $a\in[0,1]$, we will have only two free parameters for two observables, and we should be able to constrain $e$ and $f$.

\begin{figure}
 \centering
  \includegraphics[width=8.6cm]{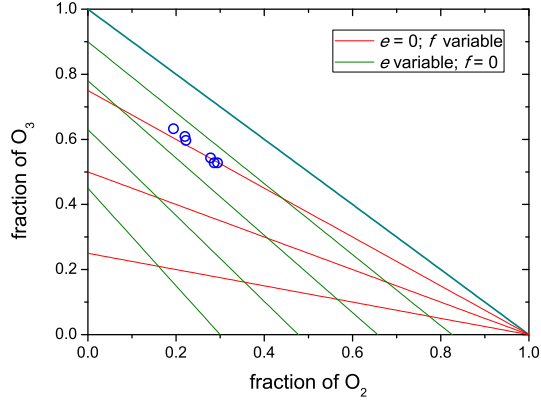}
   \caption{Results of the model of chemical desorption for values of $e$ and $f$ of 0, 0.25, 0.5, 0.75, 1. In red, the chemical desorption of (R2) is set to 0, in green the chemical desorption of (R1) is set to 0. }\label{fig:4}
    \end{figure}

Fig.~\ref{fig:4} shows the plot of $f$O$_2$(t$_f$) and $f$O$_3$(t$_f$) for $a$ varying from 0 to 1. For any given $e$ and $f$ parameters, the ensemble of possible solutions represents a straight line. In Fig.~\ref{fig:4}, keeping one parameter equal to 0, we vary the other from 0 to 1, by steps of 0.25. The case of no ozone desorption ($f$=0) is represented in green. We can see that a change of $e$ value (in green) corresponds to a shift almost parallel to the ``no chemical desorption line'' (in dark green). On the contrary, a variation of $f$ value (in red), engenders a rotation around point [1,0] and changes the slope of the straight line.

\begin{figure}
 \centering
  \includegraphics[width=8.6cm]{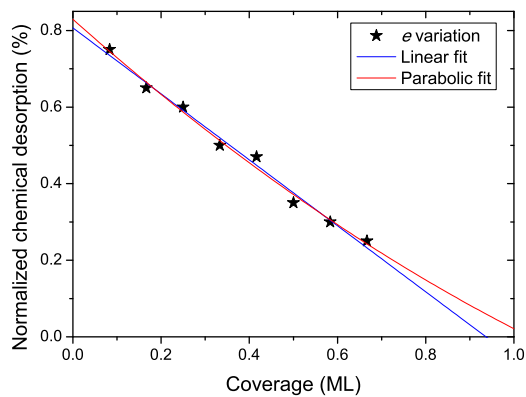}
   \caption{Black stars: variation of $e$, chemical desorption efficiency of (R1), assuming $f$ = 0, efficiency of (R2). Blue curve: linear fit assuming $e$=0 at full coverage. Red curve: parabolic fit.}\label{fig:5}
    \end{figure}

By fitting the circles points (experiments where surface temperature is varied), we find a very good matching for $e$~= 31~$\pm$~5~\% and $f$~= 5~$\pm$~4~\%. This means that most of the chemical desorption is supported by (R1).
By construction, experiments where surface temperature is varied, explore the extension of the $a$ parameter, and that is why we can constrain for the 2 other parameters $e$ and $f$. However, we want also to understand the effect of the coverage. In order to reduce the number of parameters, we decide to freeze one parameter and choose $f$~=0. This is an approximation and we will discuss later its relevance. Black stars in Fig.~\ref{fig:5} show the best $e$ parameter as function of final coverage, assuming $f=0$. We observe an almost linear decrease (blue curve in Fig.~\ref{fig:5}). Its value at the origin is 80~$\pm$1~\%. It can be prolongated to 0 for a coverage equal to unity. Actually the best fit would cross the horizontal axis at 0.94. The $e$ variation should be represented by the derivative of the data shown in Fig.~\ref{fig:5}; in fact the chemical desorption coefficient should vary with the coverage evolution, and not with the final coverage, which represents the integration of all coverages during the deposition phase. To take into account this remark, we plot a parabolic best fit of the data (red curve in Fig.~\ref{fig:5}), which is the solution of the integration of a linear function. We observe a slightly better match, but, still, the physical conclusion remains the same. The $e$ value at the origin is 0.83, and it is almost 0 (0.04) at full coverage.

We can perform the same process again by assuming a \textit{f} value of a few \%. The non zero value of \textit{f} does not affect the coordinates at the origin, due to the very low coverages, and to the consequent low ozone production. Therefore the effect on the lack of oxygen atoms is not very important and the correction at 0 coverage is small. It is about 2~\% for $f$= 5~\%. The effect on the full coverage is also a second-order effect.
Taking into account the experimental errors, due mostly to the absolute beam flux uncertainty, we can still propose that the chemical desorption efficiency in the case of (R1) is reduced proportionally with the coverage. 

\begin{figure}
 \centering
  \includegraphics[width=8.6cm]{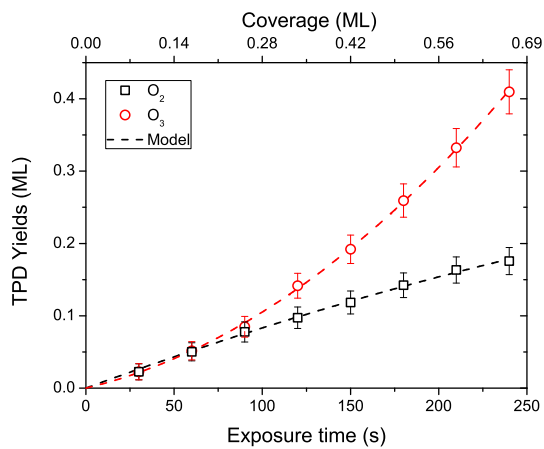}
   \caption{Black squares and red circles: O$_2$ and O$_3$ experimental yields (in ML) obtained after different exposure times (in \textit{second} in the lower x-axis, in \textit{ML} in the upper x-axis). Lines: best fit and computation of O$_2$ and O$_3$ populations, assuming linear decrease of chemical desorption with the coverage of (R1), and no chemical desorption from (R2).}\label{fig:6}
\end{figure}

Finally, we can fit our data and compute the evolution of O$_2$ and O$_3$ on the substrate, keeping the two following strong assumptions: chemical desorption is only occurring during the O+O reaction (R1), and it decreases linearly with the coverage from 80~\%  to 0 at full coverage. Fig.~\ref{fig:6} shows the comparison of the fitted results (black and red lines, respectively O$_2$ and O$_3$) with experimental points (squares and circles, respectively O$_2$ and O$_3$). The match is excellent, and the deviation can be attributed to experimental uncertainties rather than to the poor level of the simulations. We note that an excellent match can as well be obtained with the $f$ parameter frozen to few \%; however, linear dependence cannot be straightforwardly demonstrated. In fact, the accuracy of the experimental data is not sufficient to discuss within a few \%.

\section {Discussion}
From our analysis, we can claim that:
\begin{itemize}
\item[\textit{a}] the chemical desorption is mostly due to the O+O reaction with an efficiency close to 80 $\%$ (78~$\pm$5~\%) on a bare oxidized graphite. The other reaction, O+O$_2$, has a more limited reactive desorption efficiency, set to 5~$\pm$~5\%.
\item[\textit{b}] The chemical desorption is highly dependent on the presence of other molecules adsorbed on the substrate and decreases linearly with the surface coverage.
\end{itemize}
To understand these statements, we have to make clear that the two involved reactions are different in many aspects: the masses and the degrees of freedom of products, the binding energy of reactants, and the enthalpy of reaction. Tab.~I regroups the different properties of the reactions and their products.
\begin{table}[ht]
\begin{tabular}{c|c c c c|c}
  \hline
 \multirow{2}{*}{\backslashbox{Reaction}{Parameter}} & Mass & Degrees & $\Delta_f$ & E$_{binding}$ & CD            \\
                            &   (g/mol)  & of freedom & \multicolumn{2}{c|}{(kJ/mol)}   & \% \\
 \hline
R1  &    \multirow{2}{*}{32}      & \multirow{2}{*}{6} &  \multirow{2}{*}{498}   & \multirow{2}{*}{10}            & \multirow{2}{*}{79}\\
O+O $\rightarrow$O$_2$ & &&&&\\
\hline
R2  & \multirow{2}{*}{48}     & \multirow{2}{*}{9} &  \multirow{2}{*}{106}   & \multirow{2}{*}{17.5}          &  \multirow{2}{*}{5}\\
O+O$_2$ $\rightarrow$O$_3$ & &&&&\\
 \hline
\end{tabular}\label{tab:1}
\caption{List of different physical properties of reactions (R1) and (R2):  mass of reaction products expressed in \textit{g/mol}; degrees of freedom of reaction products; enthalpy of formation of each reaction expressed in \textit{kJ/mol}; binding energy of (R1) and (R2) products (O$_2$ and O$_3$, respectively) expressed in \textit{kJ/mol}.}
\end{table}

\begin{figure}
 \centering
  \includegraphics[width=8.6cm]{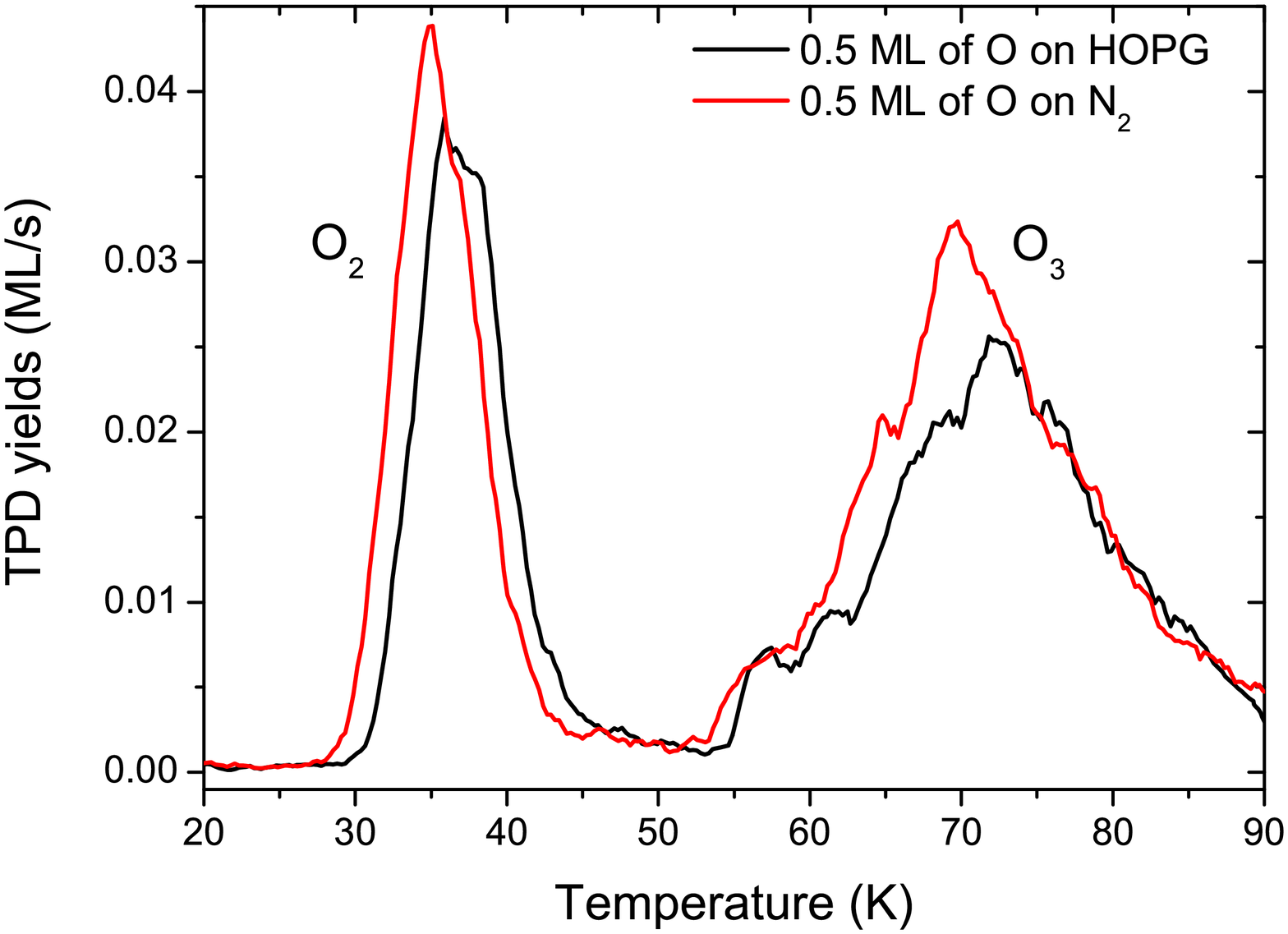}
   \caption{Black curve: O$_2$ and O$_3$ TPD traces after deposition of 0.5~ML of O+O$_2$ on a bare oxidized graphite sample. Red curve: O$_2$ and O$_3$ TPD traces after deposition of 0.5~ML of O+O$_2$ on 1~ML of N$_2$ previously deposited on a oxidized graphite.}\label{fig:7}
    \end{figure}

From Tab.~I, we can infer why chemical desorption of (R1) is larger than (R2), by looking at each of the following parameters:

\begin{enumerate}
\item \textit{Enthalpy of formation}. As already claimed, the chemical desorption process requires some energy excess: the source of the energy is certainly the enthalpy of formation, and we can see that (R1) has a larger available energy with respect to (R2).
\item \textit{Degrees of freedom}. The initial energy shall be spread among more degrees of freedom in the case of O$_3$ than in the case of O$_2$. The share of energy available for the motion perpendicular to the surface (required for desorption) is more limited in the case of O$_3$.  So the energy available for desorption favors chemical desorption of O$_2$.
\item \textit{Binding energy}. It is a limiting factor, since molecules have to overcome binding energy barrier to desorb. Here again, (R1) is favored thanks to a lower binding energy.
\item \textit{Mass of newly formed molecule}. It may have an indirect impact on the chemical desorption. Actually, phonon propagation is dependent on the mass of the colliding molecule with the surface. It is known that light molecules such as H$_2$ have a rather weak sticking coefficient (about 0.3 at room temperature) with surfaces indicating a poor energy transfer.\cite{Chaabouni12b} On the contrary heavier ones, like CO or O$_2$, have large ($>$ 0.9) sticking coefficients.\cite{Bisschop06} Of course, it depends on the type of surface, and, in some cases, collisions can be treated classically and an effective (collisional) mass of the surface can be found. The analysis of measurements of hyperthermal O$_2$ scattered  from a graphite surface shows that the effective mass of 1.8 graphite carbon ring ($\simeq130$ a.m.u.) can be adopted.\cite{Hayes12} In our case, we cannot directly use this value since we use oxidized graphite, but nonetheless we note that graphitic surfaces adopt collective properties. The collision is not made with a unique carbon atom, or eventually a pair, or even a ring, it is a collective response of the surface. We can use an effective mass somewhat larger than the mass of O$_2$ and O$_3$. To estimate the energy transfer for each molecule, we use a classic elastic collision as zero order approximation, and check the kinetic energy transfer. If the impactor has a mass $m$ and the immobile target a  mass $M$, thus the kinetic energy retained after collision for $m$ can be written as follows:
$$ \epsilon = (\frac{m-M}{M+m})^2$$
Using $m$=32 or 48 a.m.u., and $M$=130 a.m.u., we find $ \epsilon_{O_2}$ = 37$\%$ and $ \epsilon_{O_3}$ = 21$\%$ . Once again, the mass parameter favors the chemical desorption of O$_2$ molecules, because they keep more kinetic energy after the collision with the surface ($ \epsilon_{O_2}> \epsilon_{O_3}$).  
\end{enumerate}
All four parameters are in favor of R$_1$ and therefore our first experimental statement (statement \textit{a}) is fully explained. The second one (statement \textit{b}) that deals with the coverage may be explained through the fourth parameter. The energy transfer has higher values if the molecule collide with another adsorbed species: all the energy is transferred in the case of equivalent mass molecules, and 0.96 in the case of O$_3$-O$_2$ collisions. In other words, all the kinetic energy of the newly formed molecules is transferred to another adsorbed molecule upon collision. 
We have tested this scenario by adding non reactive molecules on the substrate. Fig.~\ref{fig:7} shows the TPD curves of O$_2$ and O$_3$ after depositions of 0.5~ML of O/O$_2$ respectively in a bare oxidized graphite (black curve) and on top of 1~ML of N$_2$ adsorbed on oxidized graphite (red curve). O$_2$ molecules desorb in the 25-45~K range, while ozone between 55-90~K. The red curve area is greater than the black one, which means that more O$_2$ and O$_3$ remain adsorbed on the sample at the end of reactions. This is due to a reduction of the chemical desorption efficiency. We clearly see here that energy exchange with the sample is a key parameter for the chemical desorption process. Oxidized graphite and oxidized graphite covered with N$_2$ are not the same solid samples, and as such do not produce the same results. We can note here that the chemical desorption of O$_2$ from a water substrate has been found to be almost zero.\cite{Minissale13} This is probably because the mass of water molecules allows a better momentum transfer, and as such forbids newly formed molecules to bounce on it.

Less informative, the O$_2$ peak in Fig.~\ref{fig:7} is shifted toward higher temperatures where N$_2$ is co-desorbing. This shift is due to the presence of other co-adsorbates; they can greatly change the desorption profiles as already observed in the case of isotopes of H$_2$.\cite{Dulieu05} The shift of ozone peak towards lower temperatures is due to the higher number of O$_3$ molecules adsorbed on the surface. TPD traces follow the filling behavior described in Kimmel et al or Noble et al.\cite{Kimmel01, Noble12}

\begin{figure}
 \centering
  \includegraphics[width=8.6cm]{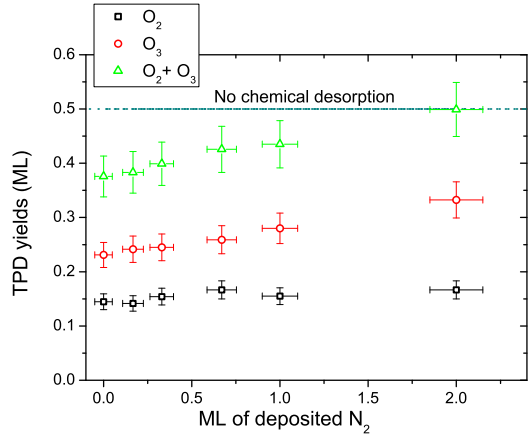}
   \caption{O$_2$ (black squares), O$_3$ (red circles) and their sum (green triangles) obtained after 0.5~ML of beam exposure in function of the dose of N$_2$ molecules previously deposited on the oxidized graphite sample. Dark green line corresponds to no chemical desorption.}\label{fig:8}
    \end{figure}

Fig.~\ref{fig:8} shows the influence of a variable amount of pre-deposited N$_2$. Total yields increase with the dose of N$_2$ pre-adsorbed on the surface. The chemical desorption process vanishes progressively and disappears between 1 and 2 pre-adsorbed layers of N$_2$. These experiments show the importance of the coverage from a different point of view. Following our previous analysis we could expect that chemical desorption vanishes exactly at one ML, due to the similar masses of O$_2$ and N$_2$ as in the case of O$_2$ and O$_3$. A small deviation from this behavior has been observed, that could be attributed to the flux calibration uncertainty, but it is more likely related to the energy transfer. The energy transfer is not only a kinetic energy transfer influenced by mass, but it is a complex ro-vibrational exchange. Previously, we have shown that the detection of excited D$_2$ molecules, ejected from an ice surface upon D recombination, disappears if the surface is before covered with a layer of  D$_2$.\cite{Congiu09} It is an evidence that internal energy of molecules is also exchanged between co-adsorbates. In the case of D$_2$, however, it is not possible to know if the molecules released in the gas phase are directly involved in reactions, or simply desorbed from thermal desorption due to surface coverage increase. In the present case, we can see if N$_2$ molecules are ejected after having absorbed energy from (R1) and (R2) reactions. This is not the case: N$_2$ desorption yield remains the same with or without Oxygen reactions. N$_2$ is clearly absorbing a part of chemical energy, although, it is not converting it efficiently into desorption. We can conclude that the indirect chemical desorption process is inefficient, within our 5~\% experimental error bars.
Moreover, desorption cannot be a direct mechanism because the initial total translational momentum is very weak (both atoms are thermalized with the substrate). Therefore, desorption happens after at least one bounce on a third body and so after a ro-vibrational and kinetic energy exchange. 
After the collision, the energy is spread again inside the molecule modes or transferred to the collider. Around 41~kJ/mol (for R1) of the energy is used for one of the dimensions of kinetic energy, if the initial chemical energy is shared equally between the degrees of freedom.  Hence, even if the energy transfer to the surface is important and only 36~\% of the energy is kept by the reactive product, there are still 15~kJ/mol of energy for (R1) to compare with 10~kJ/mol of binding energy. This can explain why the chemical desorption in case of (R1) is high on a bare and rigid surface. On the contrary, for (R2), only about 2.5~kJ/mol of energy is available to overcome the 17.5~kJ/mol energy barrier. This can explain the low chemical desorption efficiency for (R2).

In the case of the bounce on a adsorbed molecule, neither the newly formed O$_2$ molecule nor the impacted molecule have a chance to desorb: the first one transfers all the chemical energy to the second one; this last dissipates energy in its own internal degrees of freedom. By colliding with a physisorbed molecule ro-vibrational energy is shared between all the degrees of freedom of both the two molecules, and so the desorption probability is greatly reduced for both molecules.

%ICI il faut refaire la figure et bien multiplier O3 par 1.5 et tracer en gros la ligne des 0.5 ML.

Mass and energy transfer argumentations are able to depict the observed facts; nevertheless they present some limits and theoretical studies are needed to clarify how energy partition takes place. Recently, Fayolle et al\cite{Fayolle13} have studied the photo-desorption of O$_2$ using an isotopic mixture of  $^{18}$O$_2$ and $^{16}$O$_2$. They have observed  $^{16}$O$^{18}$O desorption, putting in evidence that the O+O reaction takes place. Recombinative desorption is an efficient  channel of  photo-desorption  of O$_2$, but it is not the only one. These experiments are performed with 60 layers of O$_2$, and we wonder why this mechanism does not vanish due to the high coverage like in our experiments. Actually the total yield is low, 5$\times$10$^{-3}$ molecules per photons (9~eV photons) at the maximum. We have to consider that photons are not able to induce desorption (it is only possible for the first three layers) in the deeper part of the matrix. For this reason, we can multiply by one order of magnitude, and we still obtain a very optimistic efficiency of about 5 \%. We note that our detection limit is about few $\%$ and it is compatible with such a value. Moreover, in the case of photo-desorption, O atoms formed through photon absorption have certainly important kinetic energies, due to a higher energy budget;  therefore they can react with a non negligible total momentum, on the contrary of what is presented in this paper.
Moreover, for the case of photo-desorption, the role of vibrational relaxation is central,\cite{Yuan13} which is also possible in our case, even if we
have no experimental information about it.
It is therefore difficult to compare directly experiments of  photo-desorption  and chemical desorption, even using the same studied molecule. Unfortunately in our case, we do not have any spectroscopic information and are only sensitive to overall effects. Detection of chemical desorption coupled with laser spectroscopy would provide very useful insight for the problem. It would teach us if the momentum transfer we used here is a sufficient tool to depict correctly the problem, or more certainly if detailed relaxation to others bodies (adsorbates or substrate) has to be treated for all the degrees of freedom. 
However since the details of the energy transfer at play is not yet known, our aim here was to illuminate the role of surface coverage in chemical desorption process.

\section{Conclusions}

In this paper we have shown how surface coverage influences chemical desorption process, by presenting the case of oxygen reactivity on oxidized graphite. It is a system with only two open channels of reactions leading to O$_2$ and O$_3$. Thanks to these two observables, it is possible to constrain the chemical desorption efficiency. We have shown that 
\begin{itemize}
\item reaction O+O$\rightarrow$O$_2$ is the main carrier of the oxygen loss via chemical desorption; the probability of chemical desorption is 80$\%$ on a bare surface;
\item reaction O+O$_2$ $\rightarrow$O$_3$ leads to only few percent of the gas phase products. 
\end{itemize}
We have shown that these findings are explained by referring essentially to 4 physical parameters of the reaction: (1) exothermicity of reaction, (2) mass, (3) degrees of freedom, and (4) binding energy of products.\\
Moreover, through these parameters, we are able to explain the surface coverage dependence of chemical desorption. The presence of a pre-adsorbed species enhances the probability for the newly formed excited molecules to dissipate their excess energy; in other words chemical desorption decreases with surface coverage increase. Finally, we think that the relation between the masses of newly formed species and pre-adsorbed species is a key parameter: the more similar their masses, the higher the energy transfer (lower the chemical desorption), and viceversa. As a side effect, indirect CD is unlikely.

These results represent a further step to understand how gas and solid phase are coupled in cold environments. The energy excess of the reaction between two radicals physisorbed on a cold surface can lead to the desorption of the products, even if the temperature of the substrate is low enough to keep them physisorbed. Molecular dynamic simulations will be able to investigate in more detail the energy partition of newly formed molecules.

\begin{acknowledgements}
We thank our colleagues for scientific exchange around this work: E. Congiu, H. Chaabouni, A. Moudens, S. Cazaux, G. Manic\'o, V. Pirronello and our colleagues from ISMO, H. Bergeron, D. Billy, N. Rougeau and S. Morisset.  We pay tribute to the precious technical assistance of S. Baouche.
We acknowledge the national PCMI programme funded by CNRS; the programme DIM ACAV of Region Ile de France, LASSIE, a European FP7 ITN Communitys Seventh Framework Programme under Grant Agreement No. 238258. We also thank S. Leach for a careful reading of the manuscript and helpful comments and suggestions.
\end{acknowledgements}

\end{document}